\documentclass[reprint,amsmath,amssymb,aps,prl]{revtex4-1}
\usepackage{times}
\usepackage{graphicx}
\usepackage{slashed}
\usepackage{amssymb}
\usepackage{extarrows}
\usepackage{amsmath}
\usepackage{braket}
\usepackage{xcolor}
\usepackage{units}
\usepackage{esint}
\usepackage{color}
\usepackage{lipsum}

\makeatletter
\makeatletter

\newcommand{\Yb}{\ensuremath{^{171}\mathrm{Yb}^+~}}

\begin{document}

\title{Experimental Preparation of High NOON States for Phonons}

\author{Junhua Zhang$^{1*}$, Mark Um$^{1}$, Dingshun Lv$^{1}$, Jing-Ning Zhang$^{1}$, \\ Lu-Ming Duan$^{1,2}$ and Kihwan Kim$^{1}$ }

\affiliation{
$^{1}$Center for Quantum Information, Institute for Interdisciplinary Information Sciences, Tsinghua University, Beijing, 100084, P. R. China\\
$^{2}$Department of Physics, University of Michigan, Ann Arbor, Michigan 48109, USA}

\begin{abstract} 
Multi-party entangled states have important applications in quantum metrology and quantum computation. Experimental preparation of large entangled state, in particular, the NOON states, however, remains challenging as the particle number $N$ increases. Here we develop a deterministic method to generate arbitrarily high NOON states for phonons and experimentally create the states up to $N=9$ phonons in two radial modes of a single trapped $^{171}\mathrm{Yb}^{+}~$ ion. We demonstrate that the fidelity of the NOON states are significantly above the classical limit by measuring the interference contrast and the population through the projective phonon measurement of two motional modes. We also measure the quantum Fisher information of the generated NOON state and observe the Heisenberg scaling in the lower bounds of the phase sensitivity as the $N$ increases. Our scheme is generic and applicable to other photonic or phononic systems.
\end{abstract}

\maketitle

\section*{Introduction}
Entanglement is an essential resource for quantum computation and quantum metrology. Classically, a parameter can be estimated more precisely by using more particles in the measurement, and the reduction of the statistical error is proportional to the square root of the particle number. In quantum metrology, the reduction factor can be improved to be linearly proportional to the particle number, which is called the Heisenberg limit, by using many-particle entangled states. The ultimate Heisenberg limit can be
achieved with the NOON state for identical bosons \cite{Smerzi09,Maccone11}, which can be understood by the superposition of two modes with only one of them occupied by $N$ bosons. The NOON state has the form \cite{Sanders89} 
\begin{equation}
\Ket{\psi_{{\rm NOON}}}=\frac{1}{\sqrt{2}}\left(\Ket{N,0}+e^{iN\varphi_{\rm S}}\Ket{0,N}\right),\label{eq:NOON}
\end{equation}
where the relative phase $\varphi_{\rm S}$ between two modes is linearly proportional to $N$, showing the Heisenberg scaling for parameter estimation through the interferometric measurement. For photonic systems, experiments have demonstrated NOON states with particle numbers up to $N=5$ \cite{Dowling02,Steinberg04,Zeilinger04,Takeuchi07,Dowling08,Silberberg10,Mitchell13,Xin-Yu15}. For distinguishable particles, up to $10$ photons and $14$ ions have been prepared into the closely-related GHZ states \cite{10-photon,Monz14}. NOON states have also been demonstrated in nuclear spins (NMR) \cite{Morton09}, atomic spin waves \cite{Yu-Ao10}, and microwave photons in superconducting systems \cite{Martinis11}. 

On the other hand, the quantized vibrational modes of ions in a harmonic trap have recently received increasing attention beyond the standard role as the mediator of quantum operations between internal states of ions. Phonons, bosonic quasi-particles, which represent the number of quantized excitations of a vibrational mode \cite{Meekhof96,Leibfried03}, are proposed as the information carrier for quantum simulation \cite{Porras04,Toyoda13}, Boson sampling \cite{Shen14}, and quantum computation with continuous variables \cite{Villar16}. Recently, the NOON state with $N=2$ has been generated through interference of phonons in each localized harmonic potential \cite{Kenji15}. The phonons in the trapped ion system can also be manipulated through the interaction with the internal degree of freedom of an atom, similar to manipulating photons through an atom in a cavity \cite{Raimond01}. Here, we develop a generic and deterministic scheme to generate phononic NOON states with arbitrary number of bosons $N$ for any two vibrational modes of ions based on anti-Jaynes-Cummings coupling. We experimentally generate the NOON state with phonon numbers up to $N=9$ and clearly observe the Heisenberg scaling in the lower bound of the sensitivity in the phase estimation provided by the quantum Fisher information of the state. This phononic NOON state can be applied to the precision measurement of Force or electric field gradient. 

\section*{Results}
\subsection*{Experimental Setup}
\begin{figure}[ht]
\includegraphics[width=0.45\textwidth]{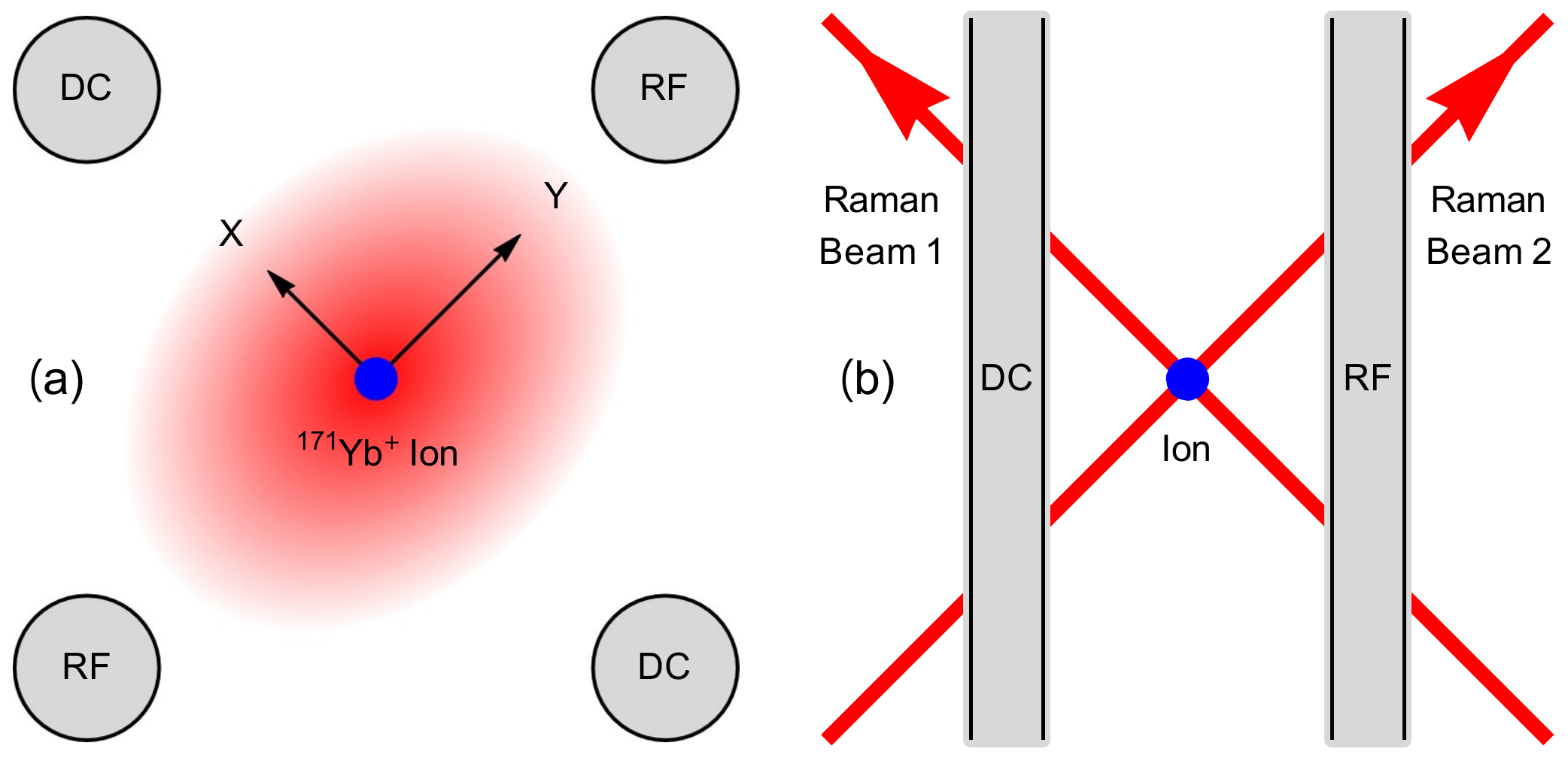} \caption{\textbf{Experimental setup.} \textbf{(a)} Side view and \textbf{(b)} top view of the
trap and laser configuration.}
\label{fig1:scheme} 
\end{figure}

In experiment, we generate the NOON state in two radial modes of an \Yb ion trapped in a standard Paul trap as shown in Fig. \ref{fig1:scheme}(a). We note that our realization is directly applicable to any normal modes of multiple ions.  
The two radial modes are denoted as X and Y, with trap frequencies
$\omega_{{\rm X}}=(2\pi)\unit[3.2]{MHz}$, $\omega_{{\rm Y}}=(2\pi)\unit[2.6]{MHz}$
and Lamb-Dicke parameters $\eta_{{\rm X}}=\Delta k\sqrt{\frac{\hbar}{2M\omega_{{\rm X}}}}=0.0538$,
$\eta_{{\rm Y}}=0.0597$, where $\Delta k$ is the difference of the
wave vector of two perpendicular Raman laser beams, $M$ is the mass
of a single \Yb ion. To mediate phonon operations, two hyperfine
levels of the \Yb ion in the $^{2}S_{1/2}$ manifold are used as
a qubit, denoted as $\Ket{\downarrow}\equiv\Ket{F=0,m_{F}=0}$ and
$\Ket{\uparrow}\equiv\Ket{F=1,m_{F}=0}$, which is separated by the
hyperfine frequency $\omega_{{\rm HF}}=(2\pi)12.6428$ GHz. The state
of the system is represented in Fock state basis as $\Ket{\sigma,n_{{\rm X}},n_{{\rm Y}}}$,
where $\sigma$ is the state of the qubit and $n_{{\rm X}}$, $n_{{\rm Y}}$
are the phonon numbers in each mode. Two laser beams from a pico-second
pulsed laser with the wavelength of 355 nm are used to generate a
stimulated Raman process to drive the carrier transition and motional
sideband transitions of the ion \cite{StiRaman}.

We operate the motional degrees of freedom with the combination of
carrier and blue-sideband pulses described by the time evolution of
the following interacting Hamiltonians of $H_{{\rm C}}$ and $H_{{\rm M}}$,
respectively \cite{Leibfried03}: 
\begin{eqnarray}
H_{{\rm C}}&=&\frac{\Omega_{{\rm C}}}{2}\left(\sigma^{+}+\sigma^{-}\right),\\ 
H_{{\rm M}}&=&\frac{i\eta_{{\rm M}}\Omega_{{\rm M}}}{2}\left(e^{i\varphi_{\mathrm{M}}}\sigma^{+}a_{{\rm M}}^{\dagger}-e^{-i\varphi_{\mathrm{M}}}\sigma^{-}a_{{\rm M}}\right),\;{\rm M=X,Y,} \nonumber \label{eq:CarBsb}
\end{eqnarray}
where $\Omega_{{\rm C}}$ and $\eta_{\mathrm{M}}\Omega_{{\rm M}}$
are the Rabi frequencies of carrier and blue sideband transitions,
$\varphi_{\mathrm{M}}$ is the phase of the driving signal, $\sigma^{+}=\Ket{\uparrow}\Bra{\downarrow}$
and $\sigma^{-}=\Ket{\downarrow}\Bra{\uparrow}$, and $a_{{\rm M}}^{\dagger}$
($a_{{\rm M}}$) is the creation (annihilation) operator of the motional
mode M (see also Methods, Hamiltonian of the System).

\subsection*{Generation Sequence}
\begin{figure}[ht]
\includegraphics[width=0.5\textwidth]{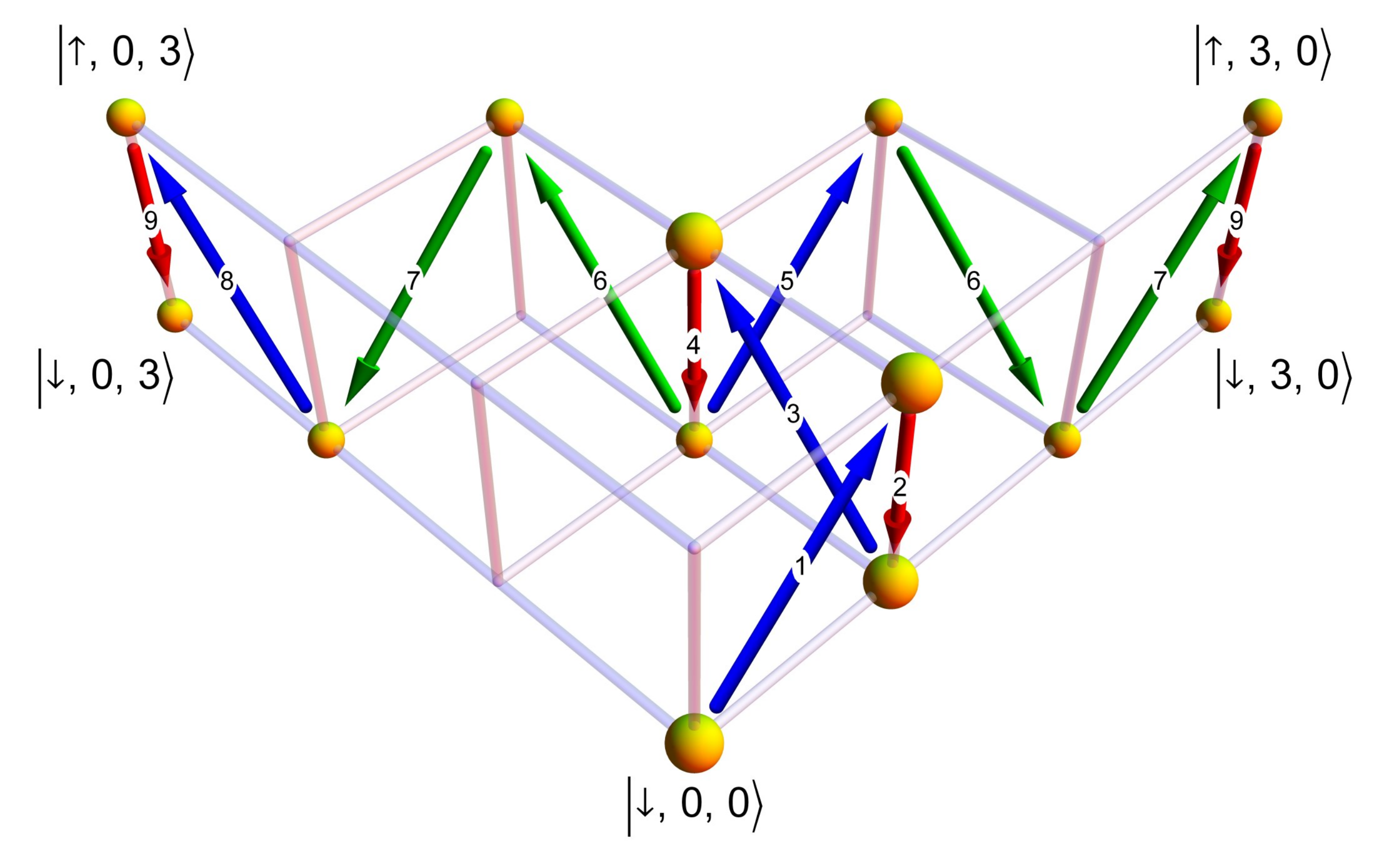} \caption{\textbf{Generation sequence of the NOON state of $\boldsymbol{N=3}$ .}
The blue arrows indicate blue-sideband transitions, the red arrows
indicate carrier transitions and the green arrows indicate composite-pulse
operations. The numbers on the arrows denote the order of the operations.}

\label{fig2:pulse} 
\end{figure}

In Fig. \ref{fig2:pulse}, we illustrate the pulse sequence for the
generation of NOON state of $N=3$, which is $\Ket{3,0}+\Ket{0,3}$,
as an example (See Methods, Pulse Sequence for a generalized description of the pulse sequence). We first initialize the state to $\ket{\downarrow,n_{{\rm X}}=0,n_{{\rm Y}}=0}$
by the standard optical pumping technique and the ground state cooling
of both motional modes using the Doppler cooling followed by the resolved
Raman-sideband cooling. Then we transfer the initial state $\Ket{\downarrow,0,0}$
to $\Ket{\downarrow,1,1}$ by applying successive $\pi$-pulses of
blue-sideband and carrier transitions. A $\frac{\pi}{2}$-pulse of
blue-sideband transition on the X mode is applied to change the state
to $\ket{\uparrow,2,1}+\Ket{\downarrow,1,1}$. Finally, two composite-pulse
operations followed by a blue-sideband $\pi$-pulse on Y mode and
a carrier $\pi$-pulse are performed to generate the state $\ket{\downarrow,3,0}+\ket{\downarrow,0,3}$.
The composite-pulse schemes are inspired by Ref. \cite{CmpPI} and
are capable of driving $\pi$-transitions of blue sideband on two
different phonon number states, which have different Rabi frequencies
(See Methods, Pulse Sequence). In order to improve the fidelity of the state, pulse-shaping
technique is applied to all blue-sideband pulses to suppress various
off-resonant couplings (See Methods, Pulse Shaping). With the pulse sequence, we
generate the NOON state up to $N=9$ which is mainly limited by experimental
imperfections that will be discussed later.  

\subsection*{Phase Sensitivity}

\begin{figure*}[ht]
\includegraphics[width=0.8\textwidth]{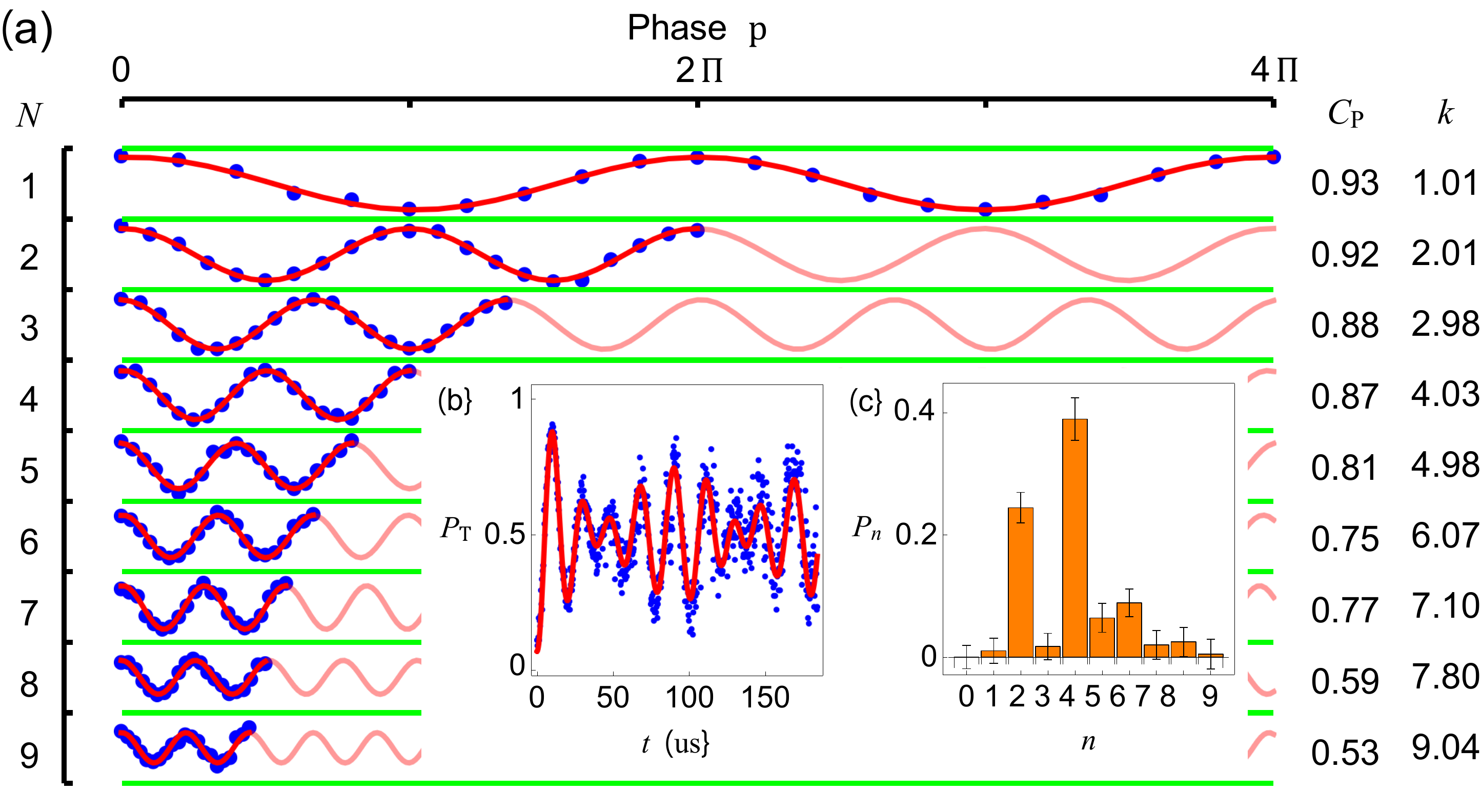} \caption{\textbf{Parity oscillations of the generated NOON states from $\boldsymbol{N=1}$
to $\boldsymbol{N=9}$.} \textbf{(a)} The blue dots are experimental data,
and the red lines are fitting curves with $\left\langle \Pi\left(\varphi\right)\right\rangle =A\cos k\varphi+B\sin k\varphi+C$,
and $C_{\mathrm{P}}=\sqrt{A^{2}+B^{2}}$. \textbf{(b)} The blue-sideband fluorescence
signal of the output mode with $\varphi=0$ for the NOON state of
$N=7$ and its fitting with $P_{\uparrow}\left(t\right)=A-\frac{1}{2}\protect\underset{n}{\sum}P_{n}\exp\left[-\left(n+1\right)^{0.7}\lambda t\right]\cos\left[\mathcal{L}_{n}^{1}\left(\eta^{2}\right)\Omega t/\sqrt{n+1}\right]$,
in which $A$, $P_{n}$, $\lambda$, $\eta$ and $\Omega$ are fitting
parameters. \textbf{(c)} The corresponding phonon distribution $P_{n}$. We
note that generally $\protect\underset{n}{\sum}P_{n}<1$ due to the
experimental errors in the generation stage. The error bars are derived
from fitting error with a confidential level of 0.95 through out the
manuscript.}

\label{fig3:parity} 
\end{figure*}

We observe the sensitivity of the phase estimation with the NOON states
increases as the number of phonons $N$ increases. The phase between
X and Y modes can be measured by the interference through the beam
splitting operation. For photons, the creation and annihilation operators
of output paths after the beam splitting operation are described by
linear combinations of those for input paths. For phonons, we can
define similar output modes written as 
\begin{eqnarray}
a_{\mathrm{O}}^{\dagger}&=&a_{\mathrm{X}}^{\dagger}\cos\theta+e^{i\varphi}a_{\mathrm{Y}}^{\dagger}\sin\theta,\nonumber \\ a_{\mathrm{O}}&=&a_{\mathrm{X}}\cos\theta+e^{-i\varphi}a_{\mathrm{Y}}\sin\theta.\label{eq:outputmode}
\end{eqnarray}
In the experiment, the parity, $\Pi=\exp\left[i\pi a_{\mathrm{O}}^{\dagger}a_{\mathrm{O}}\right]$,
of the generated state is measured in the output modes with $\theta=\pi/4$.
Depending on the value of $\varphi$, we observe the oscillation of
the parity described as 
\begin{equation}
\left\langle \Pi\left(\varphi\right)\right\rangle =C_{\mathrm{P}}\cos N\varphi.\label{eq:parity}
\end{equation}
Fig. \ref{fig3:parity}(a) shows the experimental results of the parity
oscillations from $N=1$ to $N=9$ of the generated NOON states. As
shown in the fitting parameter $k$, the enhancement of the phase
sensitivity is in agree with $N$ within 2.6\% deviation. As $N$
increases, the contrast $C_{{\rm P}}$ decreases due to experimental
imperfections. However, it is clearly shown that up to $N=9$, the
contrast is over 0.5, which indicates the existence of quantum entanglement
in the state.

The phonon distribution, and furthermore parity, of the output mode
in Eq. (\ref{eq:outputmode}) is measured through observing the time
evolution of blue-sideband transition of that mode and then fitting
the fluorescence signal \cite{Leibfried03}. The excitation of the
output mode (\ref{eq:outputmode}) is realized through driving the
blue-sideband transition of X and Y modes simultaneously shown as
\begin{widetext}
\begin{equation}
H_{\mathrm{O}}=H_{\mathrm{X}}+H_{\mathrm{Y}}=\frac{i\eta_{\mathrm{X}}\Omega_{\mathrm{X}}}{2}\left(\sigma^{+}a_{\mathrm{X}}^{\dagger}-\sigma^{-}a_{\mathrm{X}}\right)+\frac{i\eta_{\mathrm{Y}}\Omega_{\mathrm{Y}}}{2}\left(e^{i\varphi}\sigma^{+}a_{\mathrm{Y}}^{\dagger}-e^{-i\varphi}\sigma^{-}a_{\mathrm{Y}}\right).\label{eq:bsboutput1}
\end{equation}
\end{widetext}
By setting $\sqrt{2}\eta_{\mathrm{X}}\Omega_{\mathrm{X}}=\sqrt{2}\eta_{\mathrm{Y}}\Omega_{\mathrm{Y}}\equiv\Omega_{\mathrm{O}}$,
we can obtain the effective Hamiltonian for the output mode excitation
as 
\begin{eqnarray}
H_{\mathrm{O}}&=&\frac{i\Omega_{\mathrm{O}}}{2\sqrt{2}}\left[\sigma^{+}\left(a_{\mathrm{X}}^{\dagger}+e^{i\varphi}a_{\mathrm{Y}}^{\dagger}\right)-\sigma^{-}\left(a_{\mathrm{X}}+e^{-i\varphi}a_{\mathrm{Y}}\right)\right]\nonumber\\&=&\frac{i\Omega_{\mathrm{O}}}{2}\left(\sigma^{+}a_{\mathrm{O}}^{\dagger}-\sigma^{-}a_{\mathrm{O}}\right).\label{eq:bsboutput2}
\end{eqnarray}
Fig. \ref{fig3:parity}(b) shows a typical time evolution of the blue-sideband
excitation of the output mode (\ref{eq:outputmode}) with $\varphi=0$
and Fig. \ref{fig3:parity}(c) shows the phonon number distribution
by fitting the time evolution for the NOON state of $N=7$. The phase
of the generated state $\phi_{\rm S}$ is carefully measured and aligned with the
output mode of $\varphi=0$ (see Methods, Phase Alignment).

\begin{figure}[ht]
\includegraphics[width=0.45\textwidth]{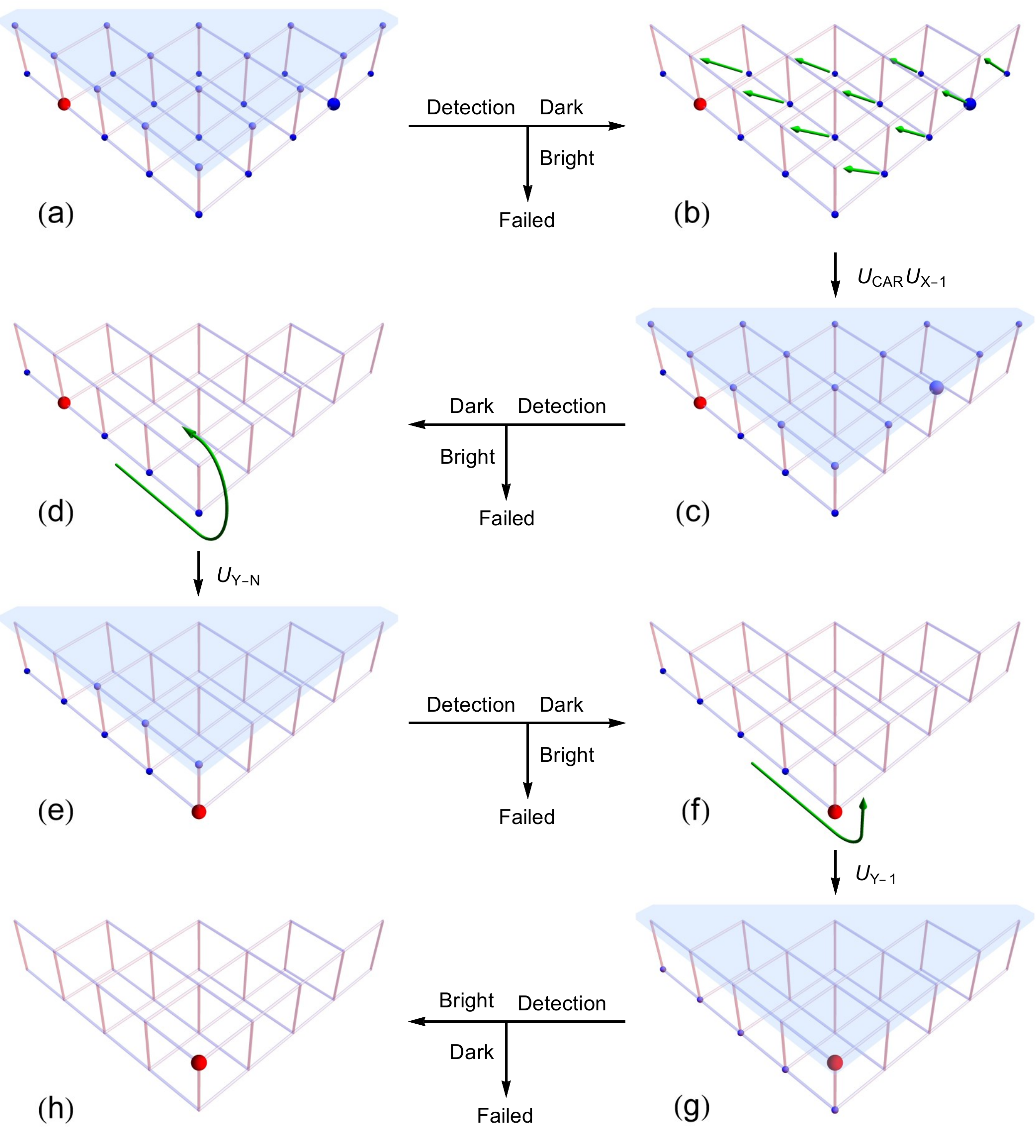} \caption{\textbf{Projective measurement of $\boldsymbol{\Ket{\downarrow,0,N}}$.}
\textbf{(a)} The state after the generation sequence. The size of the balls
indicates its amplitudes on the basis of $\Ket{\sigma,n_{{\rm X}},n_{{\rm Y}}}$.
The target state \textbf{$\Ket{\downarrow,0,N}$} of the projective
measurement is shown in red. \textbf{(b)} After the first detection stage,
the Fock states with $n_{\mathrm{X}}\geqslant1$ are transferred to
$\Ket{\uparrow}$ by operations $U_{\mathrm{X-1}}=\Ket{\uparrow,0,n_{\mathrm{Y}}}\Bra{\downarrow,0,n_{\mathrm{Y}}}+\protect\underset{n_{\mathrm{X}}>0}{\sum}\Ket{\downarrow,n_{\mathrm{X}}-1,n_{\mathrm{Y}}}\Bra{\downarrow,n_{\mathrm{X}},n_{\mathrm{Y}}}$
and then $U_{\mathrm{CAR}}=\Ket{\uparrow}\Bra{\downarrow}+\Ket{\downarrow}\Bra{\uparrow}$.
\textbf{(c)} The second detection stage. \textbf{(d)} The Fock states with $n_{\mathrm{Y}}<N$
are ``rolled'' to $\Ket{\uparrow}$ by operation $U_{\mathrm{Y-}N}=\protect\underset{n_{\mathrm{Y}}<N}{\sum}\Ket{\uparrow,n_{\mathrm{X}},N-n_{\mathrm{Y}}-1}\Bra{\downarrow,n_{\mathrm{X}},n_{\mathrm{Y}}}+\protect\underset{n_{\mathrm{Y}}\geqslant N}{\sum}\Ket{\downarrow,n_{\mathrm{X}},n_{\mathrm{Y}}-N}\Bra{\downarrow,n_{\mathrm{X}},n_{\mathrm{Y}}}$.
\textbf{(e)} The third detection stage. \textbf{(f)} Only the target state is brought
to $\Ket{\uparrow}$. \textbf{(g)} The fourth detection stage. \textbf{(h)} If the system
is projected to the target state, then fluorescence is detected in
this stage.}

\label{fig4:population} 
\end{figure}

\subsection*{Fidelity and Population Measurement}
We also measure the fidelity $F\equiv\Bra{\psi_{{\rm NOON}}}\rho_{{\rm exp}}\Ket{\psi_{{\rm NOON}}}$
of the generated NOON state. Since the density matrix of an ideal
NOON state contains only two diagonal terms and two off-diagonal terms,
the fidelity can be obtained by directly measuring these terms. The
off-diagonal terms are proportional to the contrast of the parity
oscillation $C_{\mathrm{P}}=2\left|\Bra{N,0}\rho_{\mathrm{exp}}\Ket{0,N}\right|$
(see Methods, Fidelity Analysis). For the measurement of diagonal terms, i.e., the population
of $\Ket{\downarrow,N,0}$ and $\Ket{\downarrow,0,N}$, we make use
of the arithmetic operations of phonon \cite{Um16}, which are composed
of carrier and uniform blue-sideband $\pi$-pulses.

The scheme for projective measurement of $\Ket{\downarrow,0,N}$ is
shown in Fig. \ref{fig4:population}. We first perform the fluorescence
detection, if no fluorescence occurs, the qubit state is projected
to $\Ket{\downarrow}$ {[}Fig. \ref{fig4:population}(a){]}, which
removes all the Fock states associated with $\Ket{\uparrow}$ due
to the imperfections in generating the NOON state. Then we apply one
arithmetic subtraction and a $\pi$-pulse of carrier transition, which
serves as the uniform $\pi$-transition of red-sideband in the X mode.
The operation transfers the Fock states with $n_{\mathrm{X}}\geqslant1$
from $\Ket{\downarrow}$ to $\Ket{\uparrow}$ {[}Fig. \ref{fig4:population}(b){]}.
If again no fluorescence occurs, these phonon states are eliminated
{[}Fig. \ref{fig4:population}(c){]}. Similarly for the Y mode, by
applying $N$ times of successive arithmetic subtractions and then
a detection stage, we can eliminate the Fock states with $n_{\mathrm{Y}}<N$
when no fluorescence is detected {[}Figs. \ref{fig4:population}(d)(e){]}.
We note that these operations transfer $\Ket{\downarrow,0,N}$ to
$\Ket{\downarrow,0,0}$. Finally one more subtraction operation and
the detection stage are applied {[}Figs. \ref{fig4:population}(f)(g){]}.
If the original state is projected to $\Ket{\downarrow,0,N}$, fluorescence
is observed at this detection stage. Altogether, the whole sequence
is repeated for 10,000 times and the probability of detecting fluorescence
only at the last stage of detection is the population of the $\Ket{\downarrow,0,N}$
state, $P_{0,N}$. In a similar manner, $P_{N,0}$ can be measured.

\begin{figure}[ht]
\includegraphics[width=0.5\textwidth]{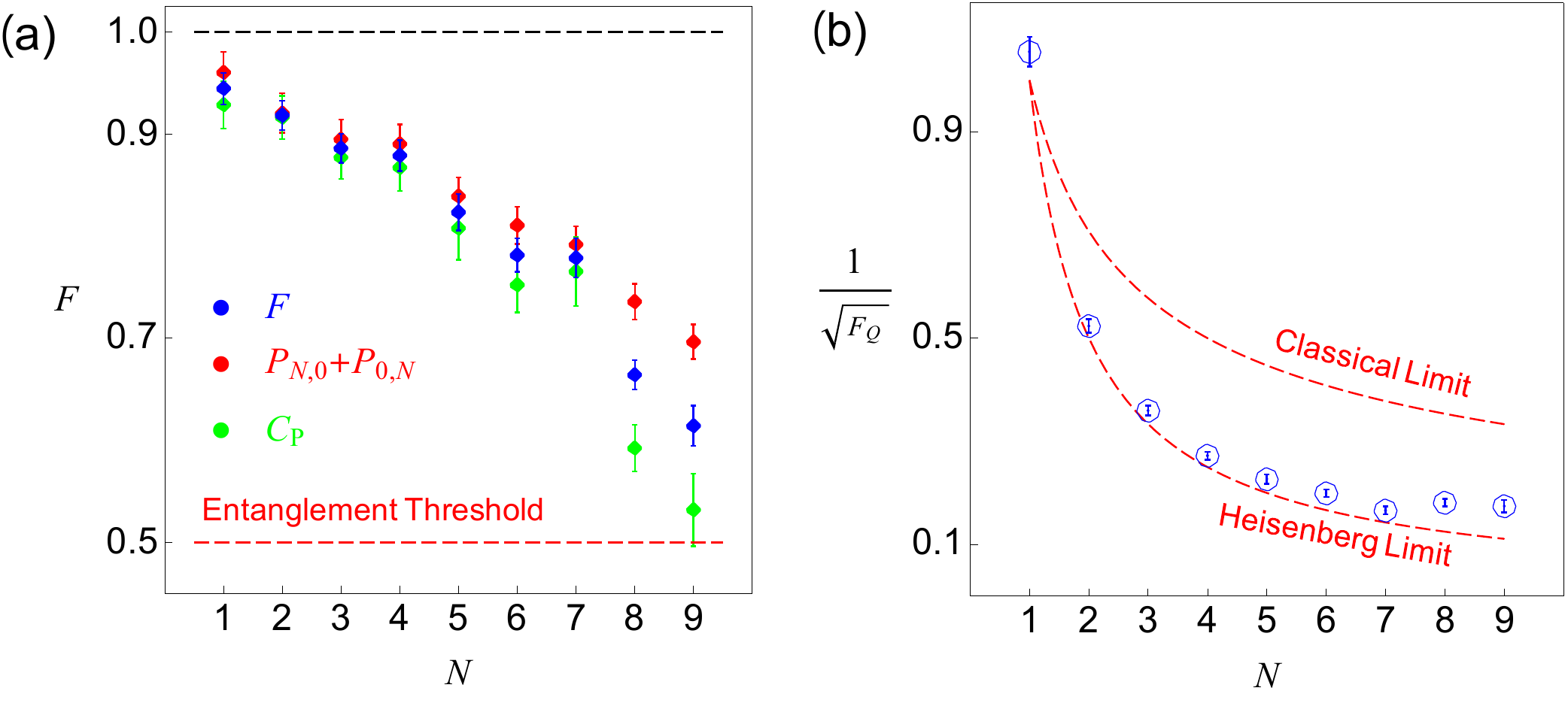} \caption{\textbf{Fidelity and quantum Fisher information of the generated states.}
\textbf{(a)} The experimental results of fidelity as well as $C_{\mathrm{P}}$
and $P_{N,0}+P_{0,N}$. The error bars of $C_{\mathrm{P}}$ are derived
from fitting error and those of $P_{N,0}+P_{0,N}$ from shot-noise
error. \textbf{(b)} The quantum Fisher information of the generated states.}

\label{fig5:fidelity} 
\end{figure}

From the results of parity and population measurements, we obtain
the fidelity (see Methods, Fidelity Analysis) of the experimental NOON state as 
\begin{equation}
F=\frac{1}{2}\left(C_{\mathrm{P}}+P_{N,0}+P_{0,N}\right).\label{eq:fidelity}
\end{equation}
As shown in Fig. \ref{fig5:fidelity}(a), the fidelities of the NOON
states up to $N=9$ are clearly larger than 0.5, which confirms these
states contain genuine multi-party entanglements.

\subsection*{Quantum Fisher Information}
Finally, we observe the Heisenberg scaling of the lower bound of the sensitivity in the
phase estimation through the quantum Fisher information (Methods, Quantum Fisher Information)
of the generated NOON states shown as 
\begin{equation}
F_{Q}=\frac{N^{2}C_{P}^{2}}{P_{N,0}+P_{0,N}}.\label{eq:fisher}
\end{equation}
The quantum Fisher information provides the best possible precision on a parameter estimation given by $1/\sqrt{F_{Q}}$ \cite{Caves94,Cooper11}, known as the Cram\'{e}r-Rao bound. For $N$ particles without entanglement, the best possible measurement scales as $1/\sqrt{N}$ and for the NOON state, the lower bound of the precision scales as $1/N$, the Heisenberg limit. As shown in Fig. \ref{fig5:fidelity}(b), the lower bound of the phase uncertainty, $1/\sqrt{F_{Q}}$, of our generated states from $N=2$ to $N=9$ clearly violate the classical bound and reach to the Heisenberg limit. 

\section*{Discussion}
This scheme of generating NOON states has no principle limit on the number of phonons $N$. Practically, various imperfections of the system prohibit the increase of the number $N$. The main problem in our system is the fluctuation of $\approx$ 10 kHz and the higher drift in the trap frequencies, which induce the increasing errors as the required number of pulses increases as $N$. The stabilization of the trap frequencies can improve the performances of pulses, which leads to the production of even higher NOON states. Our generic generation and verification scheme of the NOON states can be easily applied to any quantum system that has Jaynes-Cummings interaction including cavity or circuit QED systems \cite{Martinis11,Shi-Biao14} and optomechanical systems \cite{OptomechBgr}. We also emphasize that our realization of operating two vibrational modes through a single ion can be the essential component of large scale manipulations on multiple modes of multiple ions including boson sampling of phonons. The series of the demonstrated operations through individual ions together with the phonon number resolving detection \cite{Shuoming15,Um16} enable us to perform phononic boson-sampling. 

\section*{Methods}
\subsection*{Hamiltonian of the System}
In the experiment, we consider only the two radial modes of the ion,
so the non-interacting part of the Hamiltonian is:
\[
H_{0}=\frac{\omega_{\rm HF}}{2}\sigma_{\rm Z}+\omega_{\rm X}a_{\rm X}^{\dagger}a_{\rm X}+\omega_{\rm Y}a_{\rm Y}^{\dagger}a_{\rm Y},
\]
where $\omega_{\rm HF}$ is the frequency splitting of the qubit and $\omega_{\rm X}$, $\omega_{\rm Y}$ are the trap frequencies of two modes. We denote $\hbar\equiv1$ for convenience. 
When the ion is driven by a pair of Raman laser beams with the frequency difference $\omega$, the effective interacting Hamiltonian is written as
\begin{eqnarray*}
H_{1} & = & \Omega\cos\left(\mathbf{k}\cdot\mathbf{r}-\omega t+\phi\right)\sigma_{\rm X},\\
 & = & \frac{\Omega}{2}e^{i\left(\phi-\omega t\right)}e^{i\eta_{\rm X}\left(a_{\rm X}^{\dagger}+a_{\rm X}\right)}e^{i\eta_{\rm Y}\left(a_{\rm Y}^{\dagger}+a_{\rm Y}\right)}\sigma_{\rm X}+ {\rm h.c.}
\end{eqnarray*}
Here $\eta_{\rm X}$ and $\eta_{\rm Y}$ are the Lamb-Dicke parameters of both vibrational modes. 
For our system, $\eta_{\rm X}=0.0538,~\eta_{\rm Y}=0.0597$, it is within the Lamb-Dicke regime $N<10$ phonons	.
After taking an interaction frame with respect to $H_0$ and Lamb-Dicke approximation together with the rotating wave approximation, the Hamiltonian $H_{I}$ can be simplified as follows,
\begin{eqnarray*}
H_{\rm C}&=&\frac{\Omega}{2}\left(e^{i\phi}\sigma^{+}+e^{-i\phi}\sigma^{-}\right) \\
H_{\rm M}&=&\frac{i\eta_{\rm M}\Omega}{2}\left(e^{i\phi}a_{\rm M}^{\dagger}\sigma^{+}-e^{-i\phi}a_{\rm M}\sigma^{-}\right),
\end{eqnarray*}
$i.e.$ carrier transition $H_{\rm C}$ ($\omega=\omega_{\rm HF}$) and blue-sideband transition $H_{\rm M}$ ($\omega=\omega_{\rm HF}+\omega_{\rm M}$) for mode ${\rm M=X,Y}$, respectively. There are also red-sideband transitions for both modes
when $\omega=\omega_{\rm HF}-\omega_{\rm M}$, which are not used in the experiment.

\subsection*{Pulse Shaping}
In the theoretical analysis of the system, many off-resonant terms are neglected. However, in the experiment, these terms can severely degrade the fidelity of the generated state as the required number of pulses increases with the number of phonons. In order to achieve a higher fidelity, we implement the pulse shaping technique to suppress the effect by off-resonant couplings. The electric field of the Raman laser beams at the position of the ion with the ordinary rectangular pulses,
\[
E\left(t\right)=A\sin\left[\left(\omega-\delta\right)t+\varphi\right],
\]
is changed to a sine-shaped envelope,
\begin{eqnarray*}
E\left(t\right)&=&\frac{\pi A}{2}\sin\left[\frac{\pi t}{T}\right]\\ &\times&\sin\left[\omega t+\frac{\pi^{2}\delta}{8}\left(2\pi t-T\sin\left[\frac{2\pi t}{T}\right]\right)+\varphi\right].
\end{eqnarray*}
Here $A$ is the amplitude factor, $\omega$ and $\phi$ are the laser frequency and phase resonant to the intended transition, respectively, $\delta$ is to compensate the AC-Stark shift effect and $T$ is the duration of the pulse. First, the value of $\omega-\delta$ and the amplitude $A$ are experimentally determined with rectangular pulses and sweeping the driving frequency. The resonant frequency $\omega$ is measured by Ramsey method. And finally, the value of $\delta$ is once more carefully calibrated with a sine-shaped pulse.

\begin{table}
\centering
\scalebox{0.9}{
\begin{tabular}{|l|l|l|}
\hline 
Step & Operation & Final State\tabularnewline
\hline 
\hline 
1 & $R_{\mathrm{X}}\left(\pi/2,0,k_{\mathrm{X}}\right)$ & $\left|\uparrow,k_{\mathrm{X}}+1,k_{\mathrm{Y}}\right\rangle +\left|\downarrow,k_{\mathrm{X}},k_{\mathrm{Y}}\right\rangle $\tabularnewline
\hline 
2 & $C_{\mathrm{Y}}\left(k_{\mathrm{Y}}-1,k_{\mathrm{Y}}\right)$ & $\left|\downarrow,k_{\mathrm{X}}+1,k_{\mathrm{Y}}-1\right\rangle +\left|\uparrow,k_{\mathrm{X}},k_{\mathrm{Y}}+1\right\rangle $\tabularnewline
\hline 
3 & $C_{\mathrm{X}}\left(k_{\mathrm{X}}+1,k_{\mathrm{X}}-1\right)$ & $\left|\uparrow,k_{\mathrm{X}}+2,k_{\mathrm{Y}}-1\right\rangle +\left|\downarrow,k_{\mathrm{X}}-1,k_{\mathrm{Y}}+1\right\rangle $\tabularnewline
\hline 
4 & $C_{\mathrm{Y}}\left(k_{\mathrm{Y}}-2,k_{\mathrm{Y}}+1\right)$ & $\left|\downarrow,k_{\mathrm{X}}+2,k_{\mathrm{Y}}-2\right\rangle +\left|\uparrow,k_{\mathrm{X}}-1,k_{\mathrm{Y}}+1\right\rangle $\tabularnewline
\hline 
5 & $C_{\mathrm{X}}\left(k_{\mathrm{X}}+2,k_{\mathrm{X}}-2\right)$ & $\left|\uparrow,k_{\mathrm{X}}+3,k_{\mathrm{Y}}-2\right\rangle +\left|\downarrow,k_{\mathrm{X}}-2,k_{\mathrm{Y}}+2\right\rangle $\tabularnewline
\hline 
$\ldots$ & $\ldots$ & $\ldots$\tabularnewline
\hline 
$2k_{\mathrm{X}}$ & $C_{\mathrm{Y}}\left(k_{\mathrm{Y}}-k_{\mathrm{X}},N-2\right)$ & $\left|\downarrow,2k_{\mathrm{X}},k_{\mathrm{Y}}-k_{\mathrm{X}}\right\rangle +\left|\uparrow,1,N-1\right\rangle $\tabularnewline
\hline 
$2k_{\mathrm{X}}+1$ & $C_{\mathrm{X}}\left(2k_{\mathrm{X}},0\right)$ & $\left|\uparrow,2k_{\mathrm{X}}+1,k_{\mathrm{Y}}-k_{\mathrm{X}}\right\rangle +\left|\downarrow,0,N-1\right\rangle $\tabularnewline
\hline 
\hline 
\multicolumn{3}{|l}{For odd $N$, $k_{\mathrm{Y}}-k_{\mathrm{X}}=0$ and $2k_{\mathrm{X}}=N-1$}\tabularnewline
\hline 
$N+1$ & $R_{\mathrm{Y}}\left(\pi,0,N-1\right),R_{\mathrm{C}}$ & $\left|\downarrow,N,0\right\rangle +\left|\downarrow,0,N\right\rangle $\tabularnewline
\hline 
\hline 
\multicolumn{3}{|l}{For even $N$, $k_{\mathrm{Y}}-k_{\mathrm{X}}=1$ and $2k_{\mathrm{X}}=N-2$}\tabularnewline
\hline 
$N$ & $C_{\mathrm{Y}}\left(0,N-1\right)$ & $\left|\downarrow,N-1,0\right\rangle +\left|\uparrow,0,N\right\rangle $\tabularnewline
\hline 
$N+1$ & $R_{\mathrm{X}}\left(\pi,0,N-1\right),R_{\mathrm{C}}$ & $\left|\downarrow,N,0\right\rangle +\left|\downarrow,0,N\right\rangle $\tabularnewline
\hline 
\end{tabular}}
\caption{Pulse sequence driving $\ket{\downarrow,k_{\rm X},k_{\rm Y}}$ to NOON state $\ket{\downarrow, N,0}+\ket{\downarrow, 0,N}$}
\label{Tab1:Pulsequence}
\end{table}

\subsection*{Pulse Sequence}
In order to clearly provide a generalized description of the pulse sequence to generate the NOON state, we define the following terms for convenience: $R_{\rm C}$ denotes a carrier $\pi$-pulse and $R_{\rm X}\left(\theta,\varphi,n\right)$ denotes a blue-sideband pulse of the X mode such that the transition between $\Ket{\downarrow,n_{\rm X},n_{\rm Y}}$ and $\Ket{\uparrow,n_{\rm X}+1,n_{\rm Y}}$ has rotation angle $\theta=\sqrt{n_{\rm X}+1} \Omega t$ and $\varphi$, $R_{\rm Y}\left(\theta,\varphi,n_{\rm Y}\right)$ is similarly defined, and
\[
C_{\rm M}\left(a,b\right)\equiv R_{\rm M}\left(\pi/2,0,a\right),R_{\rm M}\left(\pi,\pi/2,b\right),R_{\rm M}\left(\pi/2,0,a\right)
\]
denotes a composite-pulse operation on mode M.
With successive blue-sideband pulses of both modes and carrier pulses, the system can be prepared to $\Ket{\downarrow, k_{\rm X}, k_{\rm Y}}$ with $k_{\rm X}=\left\lfloor \left(N-1\right)/2\right\rfloor $ and $k_{\rm Y}=\left\lfloor N/2\right\rfloor$. The remaining part of the sequence is shown in the following Table \ref{Tab1:Pulsequence}. It requires a total number of $5N-2$ pulses to generate the NOON state from $\Ket{\downarrow, 0, 0}$.

\begin{figure}[ht]
\includegraphics[width=0.45\textwidth]{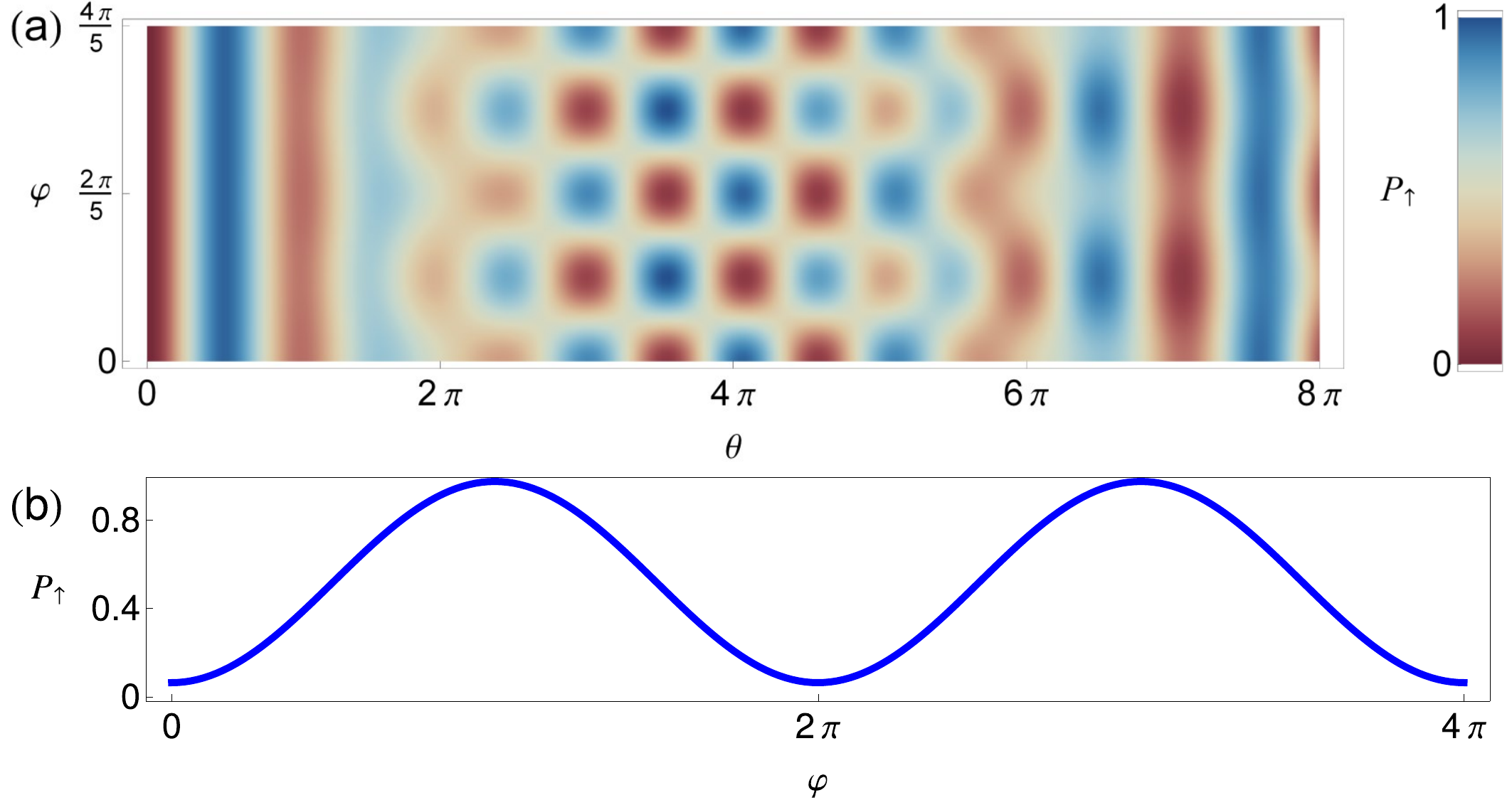} \caption{\textbf{Phase measurement for the generated NOON state of $\boldsymbol{N=5}$.}
\textbf{(a)} Theoretical calculation of the dependency of the blue-sideband fluorescence signal on $\varphi$. The duration of the excitation is measured in rotation angle $\theta=\Omega_{\mathrm{O}}t$. The red line indicates the optimal duration, $\theta=3.55\pi$, for $N=5$. \textbf{(b)} Theoretical calculation of the fluorescence signal when $\varphi$ is scanned and $\theta$ is set to the optimal. \textbf{(c)} Experimental data of a typical phase scan for $N=5$, the fitting (red line) indicates an offset of $0.15\pi$.}

\label{fig6:phasealign} 
\end{figure}

\subsection*{Phase Alignment}
We measure the phase $\varphi_{\rm S}$ of the generated NOON state and then align it to the output mode of $\varphi=0$. This phase can be measured by scanning the phase $\varphi$ of the output mode and observing the fluorescence signal of blue-sideband transition at the optimal duration [See Fig. \ref{fig6:phasealign}(a)(b)]. Fig. \ref{fig6:phasealign}(c) shows an example of the phase measurement with a result of 0.15$\pi$ for the case of $N=5$.

\subsection*{Fidelity Analysis}
We assume the density matrix of the generated state to be
\begin{eqnarray*}
\rho_{\rm exp} &=& P_{N,0} \Ket{N,0}\Bra{N,0} + P_{0,N} \Ket{0,N}\Bra{0,N} \\ &+& e^{-iN\phi} \rho_{N0,0N} \Ket{N,0}\Bra{0,N} + e^{iN\phi} \rho_{N0,0N} \Ket{0,N}\Bra{N,0} \\&+& \rho_{\rm noise},
\end{eqnarray*}
where $\rho_{\rm noise}$ stands for the irrelevant part of the density matrix and is independent of $\phi$.
The fidelity of the generated state to the ideal NOON state $\Ket{\psi_{\rm NOON}}=\frac{1}{\sqrt{2}}\left(\Ket{N,0}+e^{iN\varphi} \Ket{0,N}\right)$ is:
\begin{eqnarray*}
F & = & \Bra{\psi_{\rm NOON}}\rho_{\rm exp}\Ket{\psi_{\rm NOON}}\\
 & = & \frac{1}{2}(\Bra{N,0}\rho_{\rm exp}\Ket{N,0}+\Bra{0,N}\rho_{\rm exp}\Ket{0,N}\\
 & + & e^{iN\varphi}\Bra{N,0}\rho_{\rm exp}\Ket{0,N}+e^{-iN\varphi}\Bra{0,N}\rho_{\rm exp}\Ket{N,0})\\
 & = & \frac{1}{2}\left[P_{N,0}+P_{0,N}+e^{iN\varphi}\rho_{N0,0N}+e^{-iN\varphi}\rho_{N0,0N} \right]\\ 
 & = & \frac{1}{2}\left[P_{N,0}+P_{0,N}+2\rho_{N0,0N} \cos N\left(\varphi-\phi\right)\right].
\end{eqnarray*}
Experimentally setting the phases as $\phi=\varphi=0$, the fidelity is $F=\frac{1}{2} \left(P_{N,0}+P_{0,N}+2\rho_{N0,0N} \right).$ The values of $P_{N,0}$ and $P_{0,N}$ can be directly measured in experiment. The term $2\rho_{N0,0N}$ can be measured by the contrast $C_{P}$ of the parity oscillation of phonon observed in the output modes. In order to show the relation $2\rho_{N0,0N}\equiv C_{P}$, we introduce Schwinger\textquoteright{}s
oscillator model of angular momentum:
\begin{eqnarray*}
J_{\rm X}&=&\frac{1}{2}\left(a_{\rm X}^{\dagger}a_{\rm Y}+a_{\rm X}a_{\rm Y}^{\dagger}\right),\\
\quad J_{\rm Y}&=&\frac{1}{2i}\left(a_{\rm X}^{\dagger}a_{\rm Y}-a_{\rm X}a_{\rm Y}^{\dagger}\right),\\
\quad J_{\rm Z}&=&\frac{1}{2}\left(a_{\rm X}^{\dagger}a_{\rm X}-a_{\rm Y}^{\dagger}a_{\rm Y}\right).
\end{eqnarray*}
Then the density matrix of the system can be expressed in the angular momentum basis $\Ket{J=N/2,M_{z}}$ as:
\begin{eqnarray*}
\rho_{\rm exp} & = & P_{N,0}\Ket{J,J}\Bra{J,J}+P_{0,N}\Ket{J,-J}\Bra{J,-J}\\
 & + &\rho_{N0,0N}\Ket{J,J}\Bra{J,-J}+\rho_{N0,0N}\Ket{J,-J}\Bra{J,J}
\end{eqnarray*}

We first consider the form of the parity operator in the X mode,
\begin{eqnarray*}
\Pi&=&\exp\left[i\pi a_{\rm X}^{\dagger}a_{\rm X}\right]=\exp\left[\frac{i\pi}{2}\left(a_{\rm X}^{\dagger}a_{\rm X}-a_{\rm Y}^{\dagger}a_{\rm Y}+N\right)\right]\\ &=&\exp\left[i\pi J_{z}\right]\exp\left[i\pi J_{\rm Z}\right].
\end{eqnarray*}
With the following beam splitting operator,
\begin{eqnarray*}
U_{\rm BS}\left(\varphi\right)&=&\exp\left[-\frac{i\pi}{4}\left(a_{\rm X}^{\dagger}a_{\rm Y}e^{i\varphi}+a_{\rm X}a_{\rm Y}^{\dagger}e^{-i\varphi}\right)\right] \\ &=&\exp\left[i\pi J_{\rm X} \cos\varphi - J_{\rm Y} \sin \varphi \right],
\end{eqnarray*}
the parity operator can be transformed into the output mode as follows. 
\begin{eqnarray*}
U_{\rm BS}^{\dagger}\left(\varphi\right)\Pi U_{\rm BS}\left(\varphi\right)=e^{i\pi N}\overset{J}{\underset{M=-J}{\sum}}e^{2iM(\varphi-\pi/2)}\Ket{J,M}\Bra{J,-M}.
\end{eqnarray*}

The parity measured in the output mode is thus
\begin{eqnarray*}
\left\langle \Pi(\varphi)\right\rangle &=& {\rm Tr}\left[\rho_{\rm exp} U_{\rm BS}^{\dagger}\left(\varphi\right)\Pi  U_{\rm BS}\left(\varphi\right)\right] \\
&=& 2\rho_{N0,0N} e^{i\pi N} \cos N\left(\varphi-\pi/2\right).
\end{eqnarray*}
Therefore, the contrast of parity oscillation $C_{P}$ is thus $2\rho_{N0,0N}$.

\subsection*{Quantum Fisher Information}
In order to calculate the quantum Fisher information of the generated state, it is convenient to use the diagonal form of $\rho_{\rm exp}$,
\[
\rho_{\rm exp}=\lambda_{1}\Ket{\psi_1}\Bra{\psi_1}+\lambda_{2}\Ket{\psi_2}\Bra{\psi_2}+\rho_{\rm noise},
\]
where
\begin{gather*}
\Ket{\psi_{1}}  =  \cos\frac{\theta}{2}\Ket{N,0}+e^{iN\phi}\sin\frac{\theta}{2}\Ket{0,N}\\
\Ket{\psi_{2}}  =  \sin\frac{\theta}{2}\Ket{N,0}-e^{iN\phi}\cos\frac{\theta}{2}\Ket{0,N}\\
\rho_{\rm noise} = \underset{n>2}{\sum}\lambda_{n}\Ket{\psi_{n}}\Bra{\psi_{n}}\\
P_{N,0}+P_{0,N} = \lambda_{1}+\lambda_{2}\\
2\rho_{N0,0N} = \left|\lambda_{1}-\lambda_{2}\right|\sin\theta\equiv C_{P}.
\end{gather*}
The definition of quantum Fisher information is written as
\[
F_{Q}=\mathrm{Tr}\left[\rho\left(\phi\right)A^{2}\right]
\]
where $A$ is the symmetric logarithmic derivative operator defined by
\[
\frac{\partial\rho_{\rm exp}\left(\phi\right)}{\partial\phi}=\frac{1}{2}\left[A\rho_{\rm exp}\left(\phi\right)+\rho_{\rm exp}\left(\phi\right)A\right].
\]
With this definition, we can calculate the matrix elements of $A$ in the basis expanded by $\Ket{\psi_{i}}$
\begin{gather*}
\Bra{\psi_{i}}\frac{\partial\rho_{\rm exp}\left(\phi\right)}{\partial\phi}\Ket{\psi_{j}}=\frac{1}{2}\left(\lambda_j\Bra{\psi_{i}}A\Ket{\psi_{j}}+\lambda_i\Bra{\psi_{i}}A\Ket{\psi_{j}}\right) \\
\Bra{\psi_{i}}A\Ket{\psi_{j}}=\frac{2}{\lambda_{i}+\lambda_{j}}\Bra{\psi_{i}}\frac{\partial\rho_{\rm exp}\left(\phi\right)}{\partial\phi}\Ket{\psi_{j}}.
\end{gather*}
Note that all $\lambda_n$ and $\Ket{\psi_{n}}$ with $n>2$, which form $\rho_{\rm noise}$, are independent of $\phi$, therefore the only non-zero terms are
\[
\Bra{\psi_{1}}A\Ket{\psi_{2}}=-\Bra{\psi_{2}}A\Ket{\psi_{1}}= i\frac{\lambda_{1}-\lambda_{2}}{\lambda_{1}+\lambda_{2}} N \sin\theta.
\]

And hence:
\[
F_{Q}=\lambda_{1}\Bra{\psi_{1}} A^{2}\Ket{\psi_{1}}+\lambda_{2}\Bra{\psi_{2}} A^{2}\Ket{\psi_{2}} = \frac{N^{2}C_{P}^{2}}{P_{N,0}+P_{0,N}}
\]

\subsection*{Infidelity of the arithmetic subtraction operation}
The infidelity of the arithmetic subtraction operation, which consists
of a carrier $\pi$-pulse and a uniform blue sideband $\pi$-transition,
is evaluated in experiment as follows. The $N$ times of arithmetic addition operations,
which are just arithmetic subtraction operations in reversed order,
are applied in the X mode to drive the system from $\left|\downarrow,0,0\right\rangle $
to $\left|\downarrow,N,0\right\rangle $, then the $N$ times of
arithmetic subtraction operations are applied to bring back the state to $\left|\downarrow,0,0\right\rangle $. By detecting the probability of being in the original state $\left|\downarrow,0,0\right\rangle $, we evaluate the imperfections of the uniform transfer operations. 

We denote the total population of all $\left|\uparrow,n_{\mathrm{X}},0\right\rangle $
states as $p_{\uparrow}$, the total population of all $\left|\downarrow,n_{\mathrm{X}},0\right\rangle $
states with $n_{\mathrm{X}}>0$ as $p_{\downarrow}$ and the population
of $\left|\downarrow,0,0\right\rangle $ with $p_{0}$. The value
of $p_{\uparrow}$ is first determined by fluorscence detection immediately
after the sequence of operations. Second, by applying an extra subtraction
operation at the end of the sequence, all $\left|\uparrow,n_{\mathrm{X}},0\right\rangle $
states are transfered to $\left|\uparrow,n_{\mathrm{X}}+1,0\right\rangle $,
$\left|\downarrow,n_{\mathrm{X}},0\right\rangle $ to $\left|\downarrow,n_{\mathrm{X}}-1,0\right\rangle $
and $\left|\downarrow,0,0\right\rangle $ to $\left|\uparrow,0,0\right\rangle $,
so the value of $p_{\downarrow}$ can be determined by fluorscence
detection at this stage, and the value of $p_{0}$ is just $1-p_{\uparrow}-p_{\downarrow}$.

This test is performed with $N=5$ and $N=9$. For $N=5$, the sequence
contains 10 arithmetic operations and $p_{0}=0.797$, so the fidelity
of a single arithmetic operation is $F=0.797^{1/10}=0.9776$. And
for $N=9$, $p_{0}=0.667$, and $F=0.667^{1/18}=0.9778$. So the fidelity
of the sequence transferring $\left|\downarrow,9,0\right\rangle $
to $\left|\downarrow,0,0\right\rangle $ is 0.817. We note that the imperfection
of the operation can only decrease the detected population, which only reduces the fidelity of the generated NOON state. However, we do not recalibrate the population that surely provide the lower bound of the fidelity.

\section*{Acknowledgment}
We thank M. S. Kim, Hyunchul Nha, Chang-Woo Lee, Jeongho Bang, Su-Yong Lee, Chao Shen and Ho-Tsang Ng for the helpful suggestions and discussions. This work was supported by the National Key Research and Development Program of China under Grants No. 2016YFA0301900 (No. 2016YFA0301901), the National Natural Science Foundation of China 11374178, 11574002 and 11504197.

\end{document}